\documentstyle[12pt]{article}
\begin{document}

\begin{center}

{\large 
{\bf Non-BCS pairing in anisotropic strongly correlated electron 
systems in solids } 
                         }\\
\vskip 0.6 cm
V.~A.~Khodel$^{\dagger}$ and J.~W.~Clark\\
{\small
McDonnell Center for the Space Sciences and Department of Physics, \\
Washington University, St. Louis, MO 63130, USA \\
$^{\dagger}$ Russian Research Center, Kurchatov Institute, Moscow 123182, 
Russia 
                         } \\

\end{center}

\vskip 0.3cm
\begin{abstract}
\vskip 0.3cm
The problem of pairing in anisotropic electron systems possessing 
patches of fermion condensate in the vicinity of the van Hove points
is analyzed.  Attention is directed to opportunities for the
occurrence of non-BCS pairing correlations between the states 
belonging to the fermion condensate.  It is shown that the physical 
emergence of such pairing correlations would drastically alter the 
behavior of the single-particle Green function, the canonical pole 
of Fermi-liquid theory being replaced by a branch point.
\end{abstract}


\vskip 0.5cm
The ground state of conventional superconductors at $T=0$ is known to
be a condensate of Cooper pairs with total momentum ${\bf P}=0$.
In Fermi-liquid theory, the familiar BCS structure of the ground state
is associated with the logarithmic divergence of the particle-particle
propagator at ${\bf P}=0$ and is independent of the details of the
pairing interaction.  However, a markedly different situation can ensue
in strongly correlated systems in which the necessary stability condition
for the Landau state is violated and the Landau quasiparticle momentum
distribution suffers a rearrangement. Under certain conditions 
this rearrangement leads to a fermion condensate
(FC) -- a continuum of dispersionless single-particle (sp) states
whose energy $\epsilon({\bf p})$ coincides with the chemical potential
$\mu$ over a finite (and in general disconnected) domain ${\bf p}
\in \Omega$ in momentum space [1-11].  As a result, the preference for
pairing with ${\bf P}=0$ comes into question because of the degeneracy of
the FC sp spectrum.  In this case, the nature of pairing depends on
the configuration assumed by the FC.

Here we study a two-dimensional square-lattice system in which the
FC is situated in domains adjacent to the van Hove points, while the
sp states with ordinary dispersion are concentrated around diagonals
of the Brillouin zone \cite{vol,zkc}.  To begin with, we focus
on the nature of particle-particle correlations in the FC subsystem and
ignore contributions from the sp states with nonzero dispersion.
Traditional BCS singlet pairing correlates only the sp states
belonging to diagonally opposite patches of the FC; the description
therefore involves the single collective operator
$C_{{\bf p}}=a_{{\bf p},_-}a_{-{\bf p},_+}$ and its adjoint
$C_{\bf p}^\dagger$, which connect the ground state with states of $N$
and $N\mp 2$ particles.  However, in anisotropic electron systems
inhabiting crystalline materials manifesting fermion condensation, all
four FC patches should be treated on an equal footing.  Hence an
additional relevant collective operator
$Q_{{\bf p}}=a_{{\bf p}},_+a_{-{\bf p}+{\bf Q}},_-$,
enters the picture, together with its adjoint $Q_{\bf p}^\dagger$.
With ${\bf Q}=(\pi/l,\pi/l)$ where $l$ is the lattice constant,
this operator characterizes the pairing correlations affecting sp
states located in the neighboring FC patches.  If such additional
correlations are involved together with the ordinary BCS correlations,
the ground-state wave function evidently loses the simple BCS structure.

A salient feature of this nonabelian exemplar of the pairing problem is
the presence of two degenerate collective modes in the particle-particle
channel.  The creation of a C-pair (Cooper pair) by the operator
$C^\dagger_{\bf p}$, followed by subsequent annihilation of a Q-pair
by the operator $Q_{\bf p}$, gives rise to an excited two-particle-two-hole
state of the $N$-fermion system.  As we shall demonstrate, this process
enmeshes a whole band of many-particle-many-hole states and changes
the structure of the single-particle Green function dramatically.

We restrict considerations to the simplest, $\delta$-like form of
the interaction in the particle-particle channel, with strength
parameter $\lambda$.  Also, we assume that all the particle-hole
contributions have already been taken into account in terms of an effective
single-particle Hamiltonian having sp spectrum $\epsilon({\bf p})$
Accordingly, only pairing contributions should be incorporated in the
equation for the Green function
$G_{\alpha\beta}(x,x')=-i \langle T \psi_{\alpha}(x)\psi^\dagger_{\beta}(x')
\rangle $.
This equation, derived with the aid of equation of motion
$\left[\varepsilon-\epsilon({\bf p})\right]\psi_{\alpha}(x)-\lambda
\psi_{\gamma}^\dagger(x)\psi_{\gamma}(x )\psi_{\alpha}(x)=0$,
takes the form
\begin{equation}
(\varepsilon-\epsilon({\bf p}))G_{\alpha\beta}(x,x')+
i\lambda \langle O |T \psi_{\gamma}^\dagger(x)
\psi_{\gamma}(x\psi_{\alpha}(x)\psi_{\beta}^\dagger(x)|O \rangle
=\delta(x-x') \,.
\label{st}
\end{equation}
In the ordinary pairing problem, the average
$ \langle O| T\psi_{\gamma}^\dagger(x)\psi_{\gamma}(x)\psi_{\alpha}(x)
\psi_{\beta}^\dagger(x')|O \rangle $ is decoupled as
$ \langle  O|T a_{{\bf p},\gamma}(t)a_{-{\bf p},\alpha}(t)|C \rangle
\langle C|Ta_{{\bf p}_1,\gamma}^\dagger(t)
a_{-{\bf p}_1,\beta}^\dagger(t+\tau)|O
\rangle $.
In the generalized case being developed, where the ground state is connected
with two different states given the labels $C$ and $Q$, this same
average has the extended decomposition
$$
\langle O|Ta_{{\bf p}_1,\gamma}(t)a_{-{\bf p}_1,\alpha}(t)|C \rangle
\langle C|T a_{{\bf p},\gamma}^\dagger(t)
a_{-{\bf p},\beta}^\dagger(t+\tau)|O \rangle +
$$
$$
\qquad + \langle O|T a_{{\bf p_1},\gamma}(t)a_{-{\bf p}_1 + {\bf Q},
\alpha}(t)|Q \rangle \langle Q|T a_{{\bf p},\gamma}^\dagger (t)
a_{-{\bf p}+{\bf Q},\beta}^\dagger (t+\tau) |O \rangle  \, .
$$
For simplicity we henceforth omit spin indices $\alpha$, $\beta$, $\gamma$,
etc.  The equation for the Green function then reads
\begin{equation}
G({\bf p},\varepsilon)=G_o({\bf p},\varepsilon)
\left[(1-\Delta F_{1,0}^+({\bf p},\varepsilon)-D
F_{0,1}^+({\bf p},\varepsilon)\right] \,,
\label{bas}
\end{equation}
where
$G_o({\bf p},\varepsilon)=\left[\varepsilon-\epsilon({\bf p})\right]^{-1}$.
In the time domain, the quantity $F_{1,0}^+$ has the expression
$F_{1,0}^+({\bf p},t)  = \langle C |T a_{{\bf p}}^\dagger (t)
a_{-{\bf p}}^\dagger (t+\tau)|O \rangle $ and is interpreted as the
transition amplitude between the ground state and a state differing
from it by the presence of a single $C$-pair.  Similarly,
$ F_{0,1}^+({\bf p},\tau) =
\langle Q|T a_{{\bf p}}^\dagger (t)a_{-{\bf p}+{\bf Q}}^\dagger (t+\tau)
|O \rangle  $ is the transition amplitude between the ground state and
a state differing from it by a single $Q$-pair. The diagram-block
$\Delta \propto \langle O|Ta_{{\bf p}_1,\gamma}(t)a_{-{\bf p}_1,\alpha}(t)
|C \rangle \propto F_{1,0} (\tau=0)$ has the same meaning as
the gap order parameter of BCS theory, while the new ingredient
$D$ has the corresponding structure
$D \propto  \langle O|T a_{{\bf p_1},\gamma}(t)a_{-{\bf p}_1+{\bf Q},
\alpha}(t)|Q \rangle \propto F_{0,1}(\tau=0) $.

Employing the complementary equation of motion
$\left[\varepsilon+\epsilon({\bf p})\right] \psi_{\alpha}^\dagger (x)+
\lambda\psi_{\alpha}^\dagger (x)  \psi_{\gamma}^\dagger (x)\psi_{\gamma}(x)
=0 $, one can derive equations for the transition amplitudes $F_{1,0}^+$
and $F_{0,1}^+$:
$$
\left[\varepsilon+\epsilon({\bf p})\right]\langle C|T a_{{\bf p}}^\dagger
(t) a_{-{\bf p}}^\dagger (t+\tau)|O\rangle=- \lambda\sum_{{\bf p}_1}
\left[\langle C|T a_{{\bf p_1}}^\dagger (t)a_{-{\bf p}_1}^\dagger (t)
|O\rangle \langle O|T a_{{\bf p}}(t)a_{{\bf p}}^\dagger (t+\tau)|O\rangle +
\right.
$$
\begin{equation}
\left.
+\langle C|T a_{{\bf p}_1}^+(t)
a_{-{\bf p}_1+{\bf Q}}^+(t)|CQ^{-1}\rangle \langle CQ^{-1}|T
a_{{\bf p}}(t)a_{{\bf p}+{\bf Q}}^+(t+\tau) |O\rangle\right] \, ,
\label{f10}
\end{equation}
and analogously
$$
\left[\varepsilon+\epsilon({\bf p})\right]\langle Q|T
a_{{\bf p}}^\dagger (t) a_{-{\bf p}+ {\bf Q}}^\dagger (t{+}\tau)|O\rangle
=-\lambda\sum_{{\bf p}_1} \left[\langle Q|T a_{{\bf p}_1}^\dagger (t)
a_{-{\bf p}_1+{\bf Q}}^\dagger (t)|O\rangle
\langle O|T a_{{\bf p}}(t)a_{{\bf p}}^\dagger (t{+}\tau)|O\rangle 
\right.
$$
\begin{equation}
\left.
+ \langle Q|Ta_{{\bf p}_1}^\dagger (t)a_{-{\bf p}+1}^\dagger (t)|QC^{-1}
\rangle \langle QC^{-1}|T a_{{\bf p}}(t)a_{-{\bf p}+ {\bf Q}}^\dagger
(t+\tau)|O\rangle
\right] \,.
\label{f01}
\end{equation}

Equations (\ref{f10}) and (\ref{f01}) are conveniently rewritten 
in the compact form
\begin{eqnarray}
F_{1,0}^+&=&G_o^{-}\left[\Delta^+G+D^+F_{1,-1}\right]\  ,\nonumber\\
F_{0,1}^+&=&G_o^{-}\left[D^+G+\Delta^+F_{-1,1}\right] \  .
\label{st1}
\end{eqnarray}
where $G_o^{-}=-\left[\varepsilon+\epsilon({\bf p})\right]^{-1}$.
In contrast to BCS theory, the above system of equations for the
key quantities $G$, $F_{1,0}$, and $F_{0,1}$ is not closed:
new transition amplitudes $F_{1,-1}\sim\langle CQ^{-1}|Ta_{{\bf p}_1}(t)
a_{-{\bf p}_1+{\bf Q}}^\dagger (t+\tau)|O \rangle $ and
$F_{-1,1}\sim\langle QC^{-1}|Ta_{{\bf p}_1}(t)
a_{-{\bf p}_1+{\bf Q}}^\dagger (t+\tau)|O \rangle $
come into play, for which further equations must be derived, and so on
{\it ad infinitum}.

Upon neglecting the contributions from the new amplitudes $F_{1,-1}$
and $F_{-1,1}$, the usual BCS equations are recovered.  If we keep
these ingredients, the situation is complicated but still amenable
to analysis.  In the next step, we form equations for the amplitudes
$F_{1,-1}$ and $F_{-1,1}$ by repeating the process that led to
Eqs.~(\ref{bas}) and (\ref{st1}).  Ignoring the variation of
the blocks $\Delta$ and $D$ due changes of the states between
which the transitions occur, we have
\begin{eqnarray}
F_{1,-1}&=&-G_o\left[DF_{1,0}^++\Delta F_{2,-1}^+\right]\ ,\nonumber \\
F_{-1,1}&=&-G_o\left[\Delta F_{0,1}^++DF_{-1,2}^+\right] \,.
\label{st2}
\end{eqnarray}

Suppose now we neglect the last term on the right in each of these
equations. Then the system of equations (\ref{bas}), (\ref{st1}), and
(\ref{st2}) closes, and may be solved without difficulty.  In this
approximation, one finds
\begin{equation}
(F_{1,0}^{(2)})^+=G^{-}_o\Delta G^{(2)} / (1+K_Q) \,, \qquad
(F_{0,1}^{(2)})^+=G^{-}_oDG^{(2)}/( 1+K_C) \, ,
\end{equation}
where
\begin{equation}
K_C({\bf p},\varepsilon)=G_o({\bf p},\varepsilon)\Delta^+G_o^{-}({\bf p},
\varepsilon)\Delta \, , \qquad K_Q({\bf p},\varepsilon)=G_o^{-}({\bf p},
\varepsilon) DG_o({\bf p}+{\bf Q},\varepsilon)D^+ \,.
\end{equation}
Assembling the Green function $G$ via Eq.~(\ref{bas}), we obtain
the approximant
\begin{equation}
G^{(2)}={G_o\over Z^{(2)}}\  ,
\end{equation}
with denominator
\begin{equation}
Z^{(2)}=1+{K_C\over 1+K_Q}+{K_Q\over 1+K_C} \  .
\label{fra2}
\end{equation}

Equations for the transition amplitudes $F_{2,-1}^+$ and $F_{-1,2}^+$,
omitted in the foregoing manipulations,
may be derived along the same lines with the results
\begin{eqnarray}
F_{2,-1}^+&=&G_o^{-}\left[\Delta^+F_{1,-1}+D^+F_{2,-2}\right] \  , \nonumber\\
F_{-1,2}^+&=&G_o^{-}\left[D^+F_{-1,1}+\Delta^+F_{-2,2}\right]  \ .
\end{eqnarray}
Again deleting the last term on the right of each equation
and proceeding as before, one finds
\begin{equation}
(F_{1,0}^{(3)})^+={G^{-}_oD_C G^{(3)} \over 1+{K_Q\over 1+K_C}} \,,
\qquad (F_{0,1}^{(3)})^+={G^{-}_oD_QG^{(3)}\over 1+{K_C\over 1+K_Q}} \,,
\end{equation}
and
\begin{equation}
G^{(3)}={G_o\over Z^{(3)}} \,,
\end{equation}
with
\begin{equation}
Z^{(3)}=1+{K_C\over 1+{K_Q\over 1+K_C}}+{K_Q\over 1+{K_C\over 1+K_Q}}
\, .
\end{equation}

Indefinite continuation of this process generates the coupled set of
equations
\begin{eqnarray}
F_{n,-n}&=&-G_o\left[\Delta F^+_{n+1,-n}+DF^+_{n,-n+1}\right]\, , \nonumber\\
F_{-n,n}&=&-G_o\left[\Delta F^+_{-n+1,n}+DF^+_{-n,n+1}\right]\, , \nonumber\\
F^+_{n,-n+1}&=&G_o^{-}\left[D^+F_{n,-n}+\Delta^+F_{n-1,-n+1}\right]\,,
\nonumber\\
F^+_{-n,n+1}&=&G_o^{-}\left[D^+F_{-n,n}+\Delta^+F_{-n-1,n+1}\right]\,,
\quad {\rm etc.}
\end{eqnarray}
Together with Eq.~(\ref{bas}) for the Green function $G$, this system can be
written in closed form in terms of two new Green functions $G_C$ and
$G_Q$:
\begin{eqnarray}
G&=&G_o-G_o\Delta^+G^-_C \Delta G-G_oD^+G^-_QD G \, , \nonumber\\
G_C&=&G_o-G_o D^+G^-_QD G_C \, , \nonumber\\
G_Q&=&G_o-G_o\Delta^+G^-_C\Delta G_Q \  .
\label{eqg}
\end{eqnarray}
The system (\ref{eqg}) has the solution
\begin{equation}
G={G_o\over S+K_Q/ S} \,, \qquad
F_{1,0}={G^{-}_o\Delta G \over 1+K_Q/ S}\,, \qquad
F_{0,1}={G^{-}_oDG \over S} \,,
\label{gre}
\end{equation}
where
\begin{equation}
S=1+{G_o\Delta^+ G_o^{-}\Delta\over 1+{G_o^{-}DG_oD^+\over
1+G_o\Delta^+G_o^{-}\Delta...}}\equiv 1+{K_C\over 1+K_Q/ S} \,.
\label{aux}
\end{equation}
The latter equation, rewritten as $S^2+(K_Q-K_C-1)S-K_Q=0$, is
solved by
\begin{equation}
S={K_C-K_Q+1\over 2}+{1\over 2}\sqrt{(K_C-K_Q+1)^2+4K_Q}\  ,
\label{sol}
\end{equation}
the plus sign of the square root being chosen so to give
$S=1$ if $K_C=K_Q=0$.
With the evaluation
\begin{eqnarray}
{K_Q\over S}&=& {2K_Q\over K_C-K_Q+1+\sqrt{(K_C-K_Q+1)^2+4K_Q}}\nonumber \\
&\equiv& -{1\over 2} \left(K_C-K_Q+1- \sqrt{(K_C-K_Q+1)^2+4K_Q}\right)
\label{bau}
\end{eqnarray}
we have
\begin{equation}
1+K_Q/S=(K_Q-K_C+1)/ 2+\sqrt{(K_Q-K_C+1)^2/4+K_C} \,.
\end{equation}
It may be noted that the factors $S$ and $1+K_Q/S$ entering the formulas
for $F_{0,1}$ and $F_{1,0}$ transform into each other under interchange
of $K_C$ and $K_Q$.

Inserting relations (\ref{sol}) and (\ref{bau}) into Eq.~(\ref{gre}),
we arrive finally at
\begin{equation}
G={G_o\over{\left[(K_C-K_Q+1)^2+4K_Q\right]^{1/2}}}
\equiv {G_o\over{\left[K^2_C+K^2_Q-2K_CK_Q+2K_C+2K_Q+1\right]^{1/2}}}
\label{gref}
\end{equation}
and
\begin{equation}
F_{1,0}={G_o^-\Delta G_o\over
(1+K_Q/S){\left[(K_C-K_Q+1)^2+4K_Q\right]^{1/2}}}\,, \quad
F_{0,1}={G_o^-D G_o\over S\left[(K_C-K_Q+1)^2+4K_Q\right]^{1/2}}
\,.
\label{fref}
\end{equation}
The conventional Fermi-liquid-theory pole in the Green function
is no longer present; it has been replaced by a branch point.  This
is our primary result.  The Green function $G(\varepsilon)$ possesses
a nonzero imaginary part over a finite interval in $\epsilon$ delimited
by the two zeros of the denominator
$Z= \left[ K^2_C+K^2_Q-2K_CK_Q+2K_C+2K_Q+1 \right]^{1/2} $.
In fact, this result is generic in the sense that it is independent
of the details of the interaction in the particle-particle channel
and certainly not specific to the assumed $\delta$-function form.
Such details only affect the form of the equations for the gap
functions.

Beyond this result, there is the obvious question of whether non-BCS
pairing of the kind described here can win the contest with ordinary
BCS pairing.  To decide this issue, one needs to know the gap functions
$\Delta$ and $D$.  For our simple model interaction characterized by
a single parameter $\lambda<0$ representing the coupling strength,
these functions obey equations of the customary form
\begin{equation}
\Delta=-\lambda\int F_{1,0}({\bf p},\varepsilon)d\upsilon_{\bf p}
{d\varepsilon\over 2\pi i}\,, \qquad
D=-\lambda\int F_{0,1}({\bf p},\varepsilon)d\upsilon_{\bf p}
{d\varepsilon\over 2\pi i}\,.
\label{gaps}
\end{equation}
Here $d \upsilon_{\bf p}$ denotes the usual momentum-space volume
element, and at nonzero temperature $T$ the factor
$\tanh(\varepsilon / 2T)$ is to be inserted in the integrands.
In more realistic models, the constant $\lambda$ is replaced by
appropriate blocks ${\cal V}_{1,0}$ and ${\cal V}_{0,1}$ of Feynman
diagrams irreducible in the particle-particle channel, and the
behavior of these interactions in position or momentum space
(e.g.\ with strongly repulsive as well as attractive components)
may lead to different conclusions than those drawn below.

The solution of the gap equations (\ref{gaps}) requires a knowledge
of the sp spectrum $\xi({\bf p}) = \epsilon({\bf p})-\mu$.  As a result
of pairing correlations, $\epsilon({\bf p})$ differs from zero even
in the momentum region occupied by the fermion condensate \cite{ks1}.
The truth of this assertion becomes transparent when the BCS case
specified by $D=0$ and $\Delta=\Delta_C$ is considered.
The first of Eqs.~(\ref{gaps}) then reduces to
\begin{equation}
\Delta_C=-\lambda \int {\Delta_C\over 2E({\bf p})}d\upsilon_{\bf p}
\equiv -\lambda\int \sqrt{n({\bf p})
(1-n({\bf p}))}d\upsilon_{\bf p} \,.
\label{delc}
\end{equation}
We proceed under the assumption that pairing does not significantly affect
the momentum distribution $n_0({\bf p})$ which has been determined for
the ``normal'' system through the variational condition \cite{ks}
$\delta E_0/\delta n({\bf p})=\mu$, the presence of a FC having been
established.  Replacing $n({\bf p})$ by $n_0({\bf p})$ in Eq.~(\ref{delc}),
one obtains the relation $n_0({\bf p})=[E({\bf p})-\xi({\bf p})]/2E({\bf p})$,
where $E^2({\bf p})=\xi^2({\bf p})+\Delta^2_C$.  It follows that
$ E({\bf p})=\Delta_C/\left[4n_0({\bf p})(1-n_0({\bf p}))\right]^{1/2}$
and $\xi({\bf p})= (1-2n_0({\bf p}))\Delta/ \left[4n_0({\bf p})
(1-n_0({\bf p}))\right]^{1/2}$ \cite{ks1}.  Thus, inclusion of pairing
correlations results in an inclination of the FC plateau in the spectrum
$\xi({\bf p})$.

Next we argue that within the current model involving a single interaction
parameter $\lambda$, the energy gain produced by BCS pairing {\it exceeds}
that coming from the new pairing scenario.  To facilitate the demonstration,
we neglect the insignificant difference between
$\xi({\bf p})=\epsilon({\bf p})-\mu$
and $\xi({\bf p}+{\bf Q})=\epsilon({\bf p}+{\bf Q})-\mu$ in the FC region.
One can then verify that the system of equations (\ref{gre}) and
(\ref{gaps}) has the solution $\Delta=D\equiv \Delta_N$.  For this
case the first formula in Eq.~(\ref{gre}) simplifies to
\begin{equation}
G={G_o\over \sqrt{1+4K}} \,,
\end{equation}
and we are led to
\begin{equation}
G({\bf p},\varepsilon)={\sqrt{\varepsilon+\xi({\bf p})\over [\varepsilon-
\xi({\bf p})][\varepsilon^2-\xi^2({\bf p})-4\Delta^2_N]}}\,, \qquad
n_0(\xi)=\int\limits_{-E}^{-\xi} \sqrt{{\xi+\varepsilon\over
[\varepsilon-\xi][\xi^2+4\Delta^2_N-\varepsilon^2]}}\,
{d\varepsilon\over \pi} \,,
\end{equation}
and
\begin{equation}
\Delta_N=\lambda\int\limits_{-E}^{-\xi}{ 2\Delta_N\over
\sqrt{[\varepsilon^2-\xi^2({\bf p})] [\varepsilon^2-E^2({\bf p})}
+E^2({\bf p})-\varepsilon^2} {d\varepsilon\over 2\pi }d\upsilon_{\bf p} \,,
\end{equation}
where $E({\bf p})=\left[\xi^2({\bf p})+4\Delta^2\right]$.

To determine which of the two competitors -- BCS or non-BCS -- prevails,
we should compare the superfluid corrections $ \delta E_N$ and
$\delta E_C$ to the corresponding ground-state energies.  For this
we employ
\begin{equation}
\delta E_N=-\int_0^\lambda {\Delta^2+ D^2\over (\lambda')^2 }d\lambda'
\equiv -2 \int_0^\lambda {\Delta^2_N\over(\lambda')^2}d\lambda' \,,
\end{equation}
a result derived in the same manner as the analogous formula
for $\delta E_C$ appearing in the Landau-Lifshitz textbook \cite{lan}.
Upon further analysis it is found that at $T=0$ the gap parameter
$\Delta_N$ is markedly smaller than $\Delta_C$, which decisively
favors the BCS solution.

However, the energetic advantage of the conventional BCS solution decreases
as the temperature increases, and it disappears entirely as the critical
temperature $T_c$ is approached.  Consequently, if the coupling
constant for the states in the neighboring FC patches were to
exceed that for the states in the opposite patches, one could
encounter the same situation as in liquid $^3$H: close to $T_c$,
superfluid $^3$He is in the A-phase, but upon cooling the liquid,
the A-phase surrenders to the B-phase.  Even so, if contributions from
sp states with normal dispersion are taken into account, the conditions
for the ascendency of non-BCS pairing become more severe, suggesting
that such exotic phases are very rare visitors.

We are indebted to M. V. Zverev for numerous discussions.
This research was supported in part by NSF Grant PHY-9900713,
by the McDonnell Center for the Space Sciences, and by the
Russian Fund for Fundamental Research, Grant No.~00-15-96590 (VAK).

\end{document}